\documentclass[aps,PRL,onecolumn,groupedaddress,floatfix]{revtex4}
\usepackage{dcolumn}
\usepackage{amsmath}
\usepackage{amssymb}
\usepackage{graphicx}
\usepackage{bm}
\usepackage[T1]{fontenc}
\usepackage{color}
\usepackage{url}
\usepackage[bf]{subfigure}
\usepackage{rotating}

\begin{document}

%opening
\title{Low-dimensional behavior of Kuramoto model with inertia in complex networks}
 \author{Peng Ji$^{1,2}$}
 \author{Thomas K.DM. Peron$^3$}
 \author{Francisco A. Rodrigues$^4$} 
 \author{J\"urgen Kurths$^{1,2,5}$}
 \affiliation{$^1$Potsdam Institute for Climate Impact Research (PIK), 14473 Potsdam, Germany\\
  $^2$Department of Physics, Humboldt University, 12489 Berlin, Germany\\
$^3$Instituto de F\'{\i}sica
 de S\~{a}o Carlos, Universidade de S\~{a}o Paulo, Av. Trabalhador
 S\~{a}o Carlense 400, Caixa Postal 369, CEP 13560-970, S\~{a}o
 Carlos, S\~ao Paulo, Brazil\\
 $^4$Departamento de Matem\'{a}tica Aplicada e Estat\'{i}stica, Instituto de Ci\^{e}ncias Matem\'{a}ticas e de Computa\c{c}\~{a}o,Universidade de S\~{a}o Paulo, Caixa Postal 668,13560-970 S\~{a}o Carlos,  S\~ao Paulo, Brazil\\
 $^5$Institute for Complex Systems and Mathematical Biology, University of Aberdeen, Aberdeen AB24 3UE, United Kingdom}

\begin{abstract}
Low-dimensional behavior of large systems of globally coupled oscillators has been intensively investigated since the introduction of the Ott-Antonsen ansatz. In this report, we generalize the Ott-Antonsen ansatz to second-order Kuramoto models in complex networks. With an additional inertia term, we find a low-dimensional behavior similar to the first-order Kuramoto model, derive a self-consistent equation and seek the time-dependent derivation of the order parameter. Numerical simulations are also conducted to verify our analytical results.
\end{abstract}
\maketitle

\section*{Introduction}

Synchronization phenomena in large ensembles of coupled systems play a prominent role in many branches of natural and social sciences as well as in engineering~\cite{Pikovsky03,Arenas08:PR}. The study of collective synchronization has many applications including the modeling of the flashing of groups of fireflies~\cite{buck1988synchronous}, the collective oscillations of pancreatic beta cells~\cite{sherman1991model}, the human cardiorespiratory system~\cite{schafer1998heartbeat}, and the pedestrian induced oscillations in bridges~\cite{strogatz2005theoretical}.  A fundamental contribution to the mathematical aspects of collective synchronization was given by Kuramoto~\cite{Kuramoto_Yoshiki_lecture_notes_1975}.  In 1975 Kuramoto proposed a model to describe the behaviour of a population of coupled non-linear oscillators, employing three key simplifying assumptions~\cite{Kuramoto_Yoshiki_lecture_notes_1975}, i.e., (i) the coupling strength was chosen to be homogeneous for all pairs of coupling oscillators; (ii) the coupling strength and the natural frequency become finite; and (iii) the number of oscillators was considered to be infinite.  Diversity in the oscillators properties is usually incorporated by taking natural frequencies from a given probability distribution function. The phase transition to synchronization occurs when the coupling strength exceeds a threshold, which depends on the probability density function of the natural frequencies. 

In 2008, Ott and Antonsen~\cite{Ott_Antonsen_ansatz_chaos_2008} introduced an ansatz for studying the behaviour of globally coupled oscillators. The Ott-Antonsen ansatz has been considered to investigate continuously time-dependent collective behavior~\cite{PhysRevE.86.046212} and for the study of delay heterogeneity~\cite{PhysRevLett.103.044101}. In addition, such ansatz has enabled to find nonuniversal transitions to synchrony in the model with a phase lag for certain unimodal frequency distributions~\cite{PhysRevLett.109.164101}. 

Although these works have provided important contributions to synchronization theory, only oscillators with global coupling have been taken into account~\cite{PhysRevE.86.046212,PhysRevLett.103.044101,PhysRevLett.109.164101,PhysRevLett.110.064101,PhysRevE.88.012905}.  Thus, a natural extension of these works can investigate how these results change when different coupling schemes are introduced. Barlev et al.~\cite{Barlev011}  studied the dynamics of coupled phase oscillators, but such approach involved integrating $N$ ordinary differential equations. To overcome this limitation, in this report we generalize the Ott-Antonsen ansatz to complex networks in the continuum limit to investigate a time-dependent phase transition to synchronization. We reduce the dimension of the system of equations from $N$ to the number of possible degrees in the network. 

Motivated by the results of the first-order Kuramoto model, we substantially extend the theory to the second-order Kuramoto model.
The Kuramoto model with inertia has been widely used for deepening the understanding of power grids~\cite{PhysRevLett.110.218701,strogatz1994nonlinear,Acebron05:RMP,Dorfler14012013}, superconducting Josephson Junctions~\cite{strogatz1994nonlinear} and many other applications~\cite{strogatz1994nonlinear,tanaka1997self}. Therefore a theory that investigates the low-dimensional character of such systems giving access to their time-dependent behavior can bring important new insights into the study. We substantially address this problem for what is perhaps the simplest choice of inertia term. In this case, the Fourier series expansion, the key approach of the Ott-Antonsen ansatz, no longer applies directly. Thus, a generalized framework for the second derivative needs to be developed, as already pointed out in recent studies~\cite{PhysRevE.88.012905,sonnenschein2013approximate}. In order to fill this gap, we derive self-consistent equations and seek the time evolution of the order parameter. Comparison of analytical and simulation results shows a good agreement. Our results shed light on the impact of the topology on the global dynamics.

\section*{Results}

We consider the first-order Kuramoto model on an unweighted and undirected complex network. The state of oscillator $i$ is denoted by its phase $\theta_i  (i=1, 2, \cdots, N)$, and the governing equation of the model~\cite{Kuramoto_Yoshiki_lecture_notes_1975} is 
\begin{equation}
\frac{d\theta_i}{dt}=\Omega_i+K\sum^{N}_{j=1} A_{ij}\sin(\theta_j-\theta_i),
\label{First_order_KM_1}
\end{equation}
where $\Omega_i$ stands for the natural frequency of oscillator $i$, which is distributed according to some probability density $g(\Omega)$, $K$ specifies the homogeneous coupling strength between interconnected nodes, and $A_{ij}$ is the element of the adjacency matrix $\textbf{A}$, i.e., $A_{ij}=1$ if nodes $i$ and $j$ are connected or $A_{ij}=0$, otherwise.

In uncorrelated networks, if $N$ approaches infinity (in thermodynamic limit), the probability of  selecting an edge connected to a node with degree $k$, natural frequency $\Omega$, and phase $\theta$ at time $t$ is $kP(k)\rho(k;\Omega,\theta,t)$, where we define $P(k)$ as the degree distribution and $\rho(k;\Omega,\theta,t)$ as the probability distribution function of nodes with degree $k$ that have natural frequency $\Omega$ and phase $\theta$ at time $t$~\cite{Ichinomiya04:PRE,PhysRevE.86.056108,PhysRevLett.110.218701}.

To characterize the macroscopic behavior of the oscillators, in the continuum limit, we consider the order parameter (see Methods for details) 
\begin{eqnarray}
re^{i\psi} &=& \int dk \int d\Omega \int d\theta P(k)k\rho(k;\Omega,\theta,t)e^{i\theta}/\int dk P(k)k\nonumber\\
&=&\int dk P(k)k r_ke^{i\psi_k} /\int dk P(k)k,
\label{order_parameter_global}
\end{eqnarray}
where $r_k$ quantifies the local synchrony of oscillators with degree $k$
\begin{equation}
r_ke^{i\psi_k}=\int d\Omega \int d\theta \rho(k;\Omega,\theta,t)e^{i\theta}.
\label{order_parameter_local}
\end{equation} 

For simplicity, we assume that the natural frequencies $\Omega_i$ are distributed according to an unimodal and symmetric Cauchy-Lorentz distribution ($g(\Omega)$) (see Methods for details) with zero mean. We set $\psi=\psi_k=0$ without loss of generality~\cite{strogatz2000kuramoto}. The coupling term in Eq.~(\ref{First_order_KM_1}) can be written as
$
	\sum^{N}_{j=1} A_{ij}\sin(\theta_j-\theta_i)=k_ir \mathrm{Im}[e^{\psi-\theta_i}]
$~\cite{Ichinomiya04:PRE,PhysRevE.86.056108,PhysRevLett.110.218701}.
Thus the governing equation~(\ref{First_order_KM_1}) can be rewritten as
\begin{equation}
\frac{d\theta}{dt}=\Omega+Kkr\frac{e^{-i\theta}-e^{i\theta}}{2i},
\label{First_order_KM_2}
\end{equation}
 which shows that the oscillators are coupled via the mean-field order parameter $r$. The restoring force tends to bring each oscillator towards equilibrium and the amount of forcing is proportional to its degree $k$. 
 
The evolution of $\rho(k;\Omega,\theta,t)$ is governed by the continuity equation, i.e.,
$
\frac{\partial \rho}{\partial t} +\frac{\partial \rho v}{\partial \theta}=0,
$
where 
$
v(k;\Omega,\theta,t)=\frac{d\theta}{dt}.
$
We use the Ott-Antonsen ansatz~\cite{Ott_Antonsen_ansatz_chaos_2008} and expand the density function in a Fourier series, i.e., 
\begin{equation}
\rho(k;\Omega,\theta,t)=\frac{g(\Omega)}{2\pi}\left\{ 1+\left[\sum\limits _{n=1}^{\infty}[a(k;\Omega,t)]^{n}e^{in\theta}+c.c.\right]\right\}.
\label{OA_ansatz_Fourier}
\end{equation}
where $c.c$ stands for the complex conjugate. Substituting the expansion into Eqs.~(\ref{order_parameter_local}) and in the continuity equation, we get that $r_k=a(k)$ and $r_k$ evolve according to 
\begin{equation}
\frac{dr_{k}}{dt}=-r_{k}+\frac{Kkr}{2}(1-r_{k}^{2})\;\;\mbox{for $k \in [k_{\min},k_{\max} ]$}
\label{r_k_evolves}
\end{equation}
where $k_{\min}$ and $k_{\max}$ are the minimum and the maximum degree, respectively. This method works efficiently compared to~\cite{Barlev_2011_chaos_an_ensemble_approach} especially when the power law behavior has some cutoff~\cite{PhysRevLett.107.178701}.
$a(k)$ therein allows a clear physical interpretation as measuring the internal synchrony of the nodes with the same degree $k$. 
The global order parameter $r$ is a sum of different $r_k$ multiplied by their degree and degree distribution (see Eq.~(\ref{order_parameter_global})).  

\begin{figure}
\includegraphics[width=0.9\linewidth]{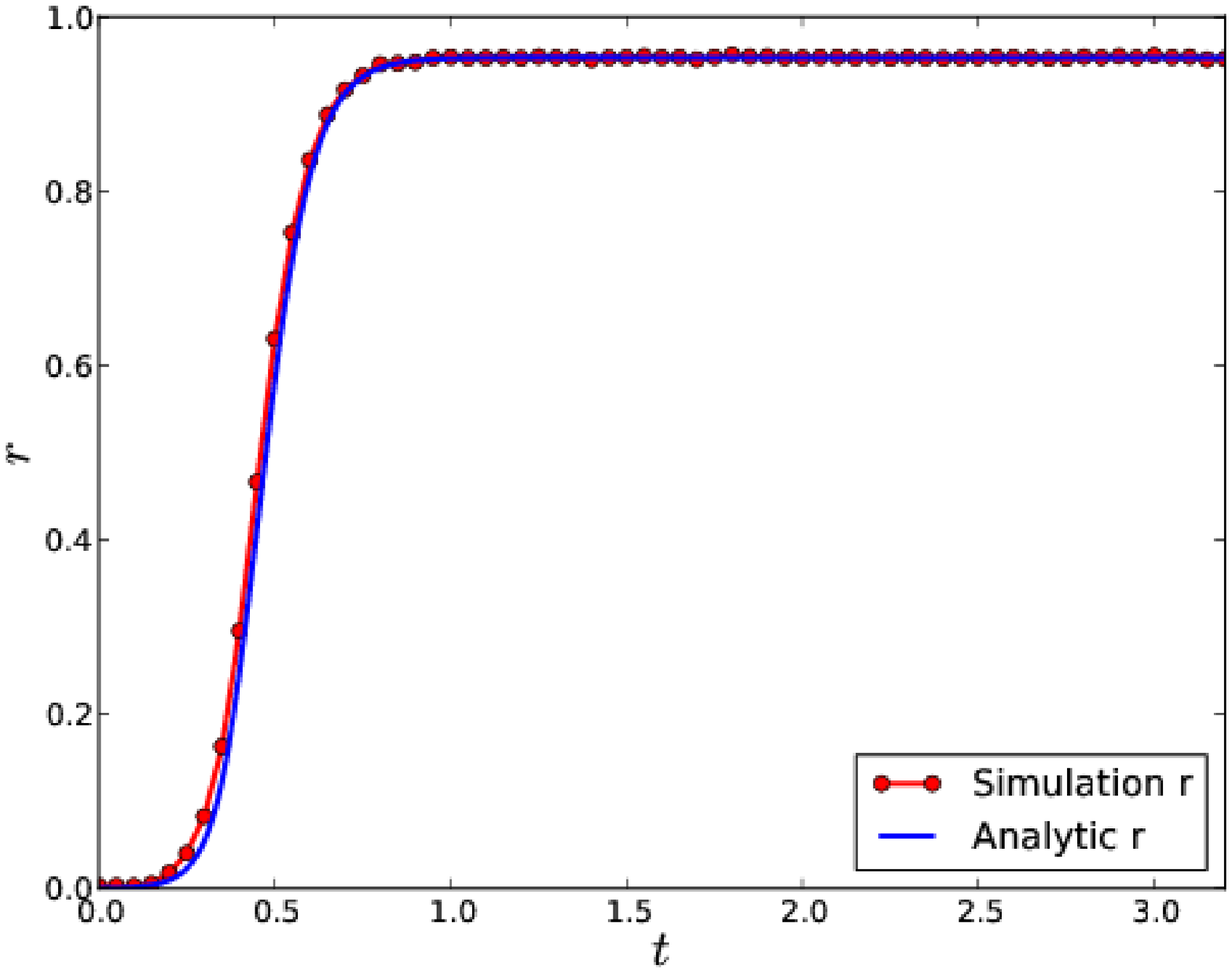}
\caption{(color online). The order parameter as a function of time. Numerical simulations of the Kuramoto model are conducted on a scale-free network (see Methods for details). The coupling strength $K=2.5$ and $\theta$ are randomly selected from $-\pi$ to $\pi$ at $t=0$.}
\label{Fig:figure_r_1st}
\end{figure}

To verify the accuracy of the time evolution of the order parameter $r_k$ (see Eq.~(\ref{r_k_evolves})), we compare the time evolution of the order parameter $r$ with numerical simulations. Fig.~\ref{Fig:figure_r_1st} shows the results. Initially, the values of oscillators are selected at random from $\pi$ to $-\pi$, which implies that the initial value of each $r_k(0)$ tends to zero.  In our simulations, we set $r_k(0)=0.001$. As we can see in Fig.~\ref{Fig:figure_r_1st}, the results obtained through the solution of the reduced system in Eq.~\ref{r_k_evolves} are in good agreement with the numerical simulations.

The analysis above shows the remarkable usefulness of the Ott-Antonsen ansatz of the first-order Kuramoto model in complex networks, but what happens when we consider the Kuramoto model with inertia? The simplest and most straightforward way is to include one unity inertia term. This leads to the mean-field character of the second-order Kuramoto model
~\cite{PhysRevLett.110.218701,strogatz1994nonlinear,Acebron05:RMP,Dorfler14012013}
\begin{equation}
\frac{d^2\theta_i}{dt^2}=-\frac{d\theta_i}{dt}+\Omega_i+Kkr\sin(-\theta_i),
\end{equation}
where $k$ varies from the minimal to the maximal degree.

As shown in Eq.~(\ref{OA_ansatz_Fourier}), the main idea of the Ott-Antonsen ansatz is to expand the probability density $\rho(k;\Omega,\theta,t)$ in a Fourier series in $\theta$. For the Kuramoto model with inertia, the probability density $\rho(k;\Omega,\theta,\dot{\theta},t)$ is also a function of the additional term $\dot{\theta}$. As $\dot{\theta}$ varies from $-\infty$ to $\infty$, it is not possible to follow the same procedure to derive the nonlinear evolution of the order parameter $r$. Due to the existence of the inertia term and the bistable area of the stability diagram~\cite{strogatz1994nonlinear}, we rewrite Eq.~(\ref{r_k_evolves}) with two functions $\Lambda(Kk)$ and $f(Kk,r) \equiv a(Kk)^br^c$ and get    
\begin{equation}
\frac{dr_k}{dt}= -r_k+\Lambda(Kk)\frac{r(1-r_k^2)}{2} + f(Kk,r_k),
\label{seek_nonlinear_time_evolution_second}
\end{equation}
where $\Lambda(Kk)$ indicates the effective coupling strength and $a$, $b$ and $c$ are constant. $f(Kk,r)$ is a high-order term and is used to adjust the stationary solution. For the Kuramoto model without inertia, we get $\Lambda(Kk)=Kk$ and $a=0$.

In order to solve Eq.~(\ref{seek_nonlinear_time_evolution_second}), we first investigate the nonlinear dynamics on fully connected networks. In this case, we normalize the coupling strength from $KN$ to $K$.
For the sake of convenience, we change the time scale to $\tau=\sqrt{Kr}t$, which yields
\begin{equation}
\frac{d^2\theta_i}{d\tau^2}=-\beta\frac{d\theta_i}{d\tau}+I_i+\sin(-\theta_i),
\end{equation}
where 
$\beta\equiv1/\sqrt{Kr}$ and $I_i\equiv\Omega_i/(Kr)$. Thus $\beta$ is identical for all oscillators and the diversity of $I_i$ is due to its natural frequency. According to the parameter space~\cite{strogatz1994nonlinear,PhysRevLett.110.218701}, nodes are divided into three groups. Melnikov's method~\cite{guckenheimer1983nonlinear} is used to show that oscillators are within a stable fixed point area as $\beta\rightarrow0$ and $I\leq 4\beta/\pi$; only limit-cycle oscillators exist for $I>1$; limit cycles and stable fixed points coexist otherwise.

\begin{figure}
\includegraphics[width=0.9\linewidth]{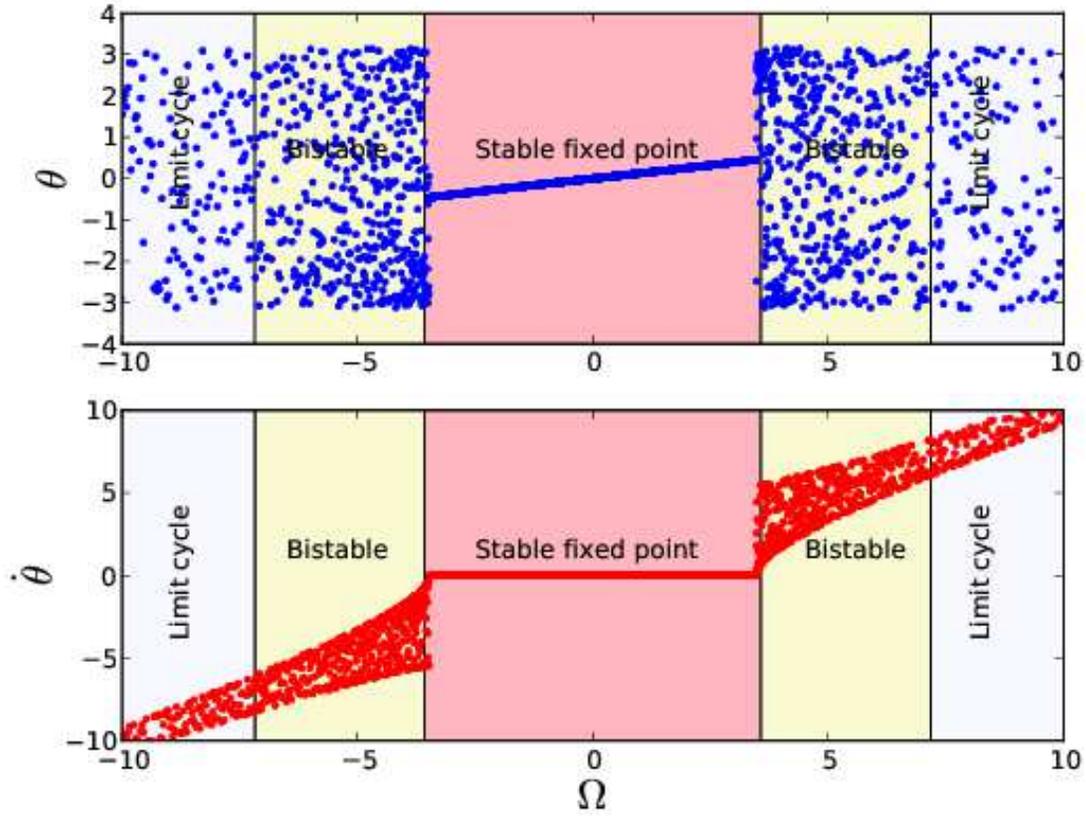}
%\caption{Phase_VS_natural_frequency_stationary_states}
\caption{(color online). Phases $\theta$ and frequencies $\dot{\theta}$ vs natural frequencies $\Omega$, which shows that phase-locked oscillators only exist in red area but not in the yellow area. The read area indicates parameter combination of stable fixed point. Stable fixed points and limit cycles coexist in the yellow area. The white area represents the existence of limit cycles. The stationary value of the order parameter $r$ could be calculated by simulations or Eq.~(\ref{r_self_consisent}). Thus nodes with natural frequencies between $[-4\sqrt{Kr}/\pi,4\sqrt{Kr}/\pi]= [-3.57,3.57]$ are synchronized. The boundary of bistable region are specified by $|\Omega|$ within $[4\sqrt{Kr}/\pi,Kr]=[3.57,7.18]$.} 
\label{Fig:natural_frequency_theta}
\end{figure}

Let us first investigate the stationary states of phases $\theta$ and $\dot{\theta}$ in terms of the natural frequencies $\Omega$ separately. In Fig.~\ref{Fig:natural_frequency_theta}, every single point represents the state of one oscillator at time $T$ $(T>>1)$ using simulations with $N=10000$ nodes and degree $K=10$. It is interesting to find that instead of three different regions mentioned above, the oscillators fall into either of the following two groups. (i) If the natural frequencies of nodes are within the boundary of the phase synchronization regime $[\Omega_{\textrm{lower}},\Omega_{\textrm{upper}}]\equiv[-4\sqrt{Kr}/\pi,4\sqrt{Kr}/\pi]$ which is the same as the above stable fixed points area, these nodes converge to fixed points and the stationary state of phases are functions of $\Omega$, which are equal to $\arcsin(\Omega/(Kr))$. This boundary is smaller than that of the Kuramoto model, in which oscillators are in the locked state for all $|\Omega| \leq Kr$~\cite{strogatz2000kuramoto}. (ii) In contrast, the oscillators with $ |\Omega| > 4\sqrt{Kr} / \pi$ are drift. 
Thus, in networks, instead of three different areas of single pendulum model, only two distinct areas could exist: fixed point and limit cycle. Nodes with the same natural frequency are either converging to single fixed points or oscillating periodically; and nodes always return to previous states even after large perturbations. 

\begin{figure}
\includegraphics[width=0.9\linewidth]{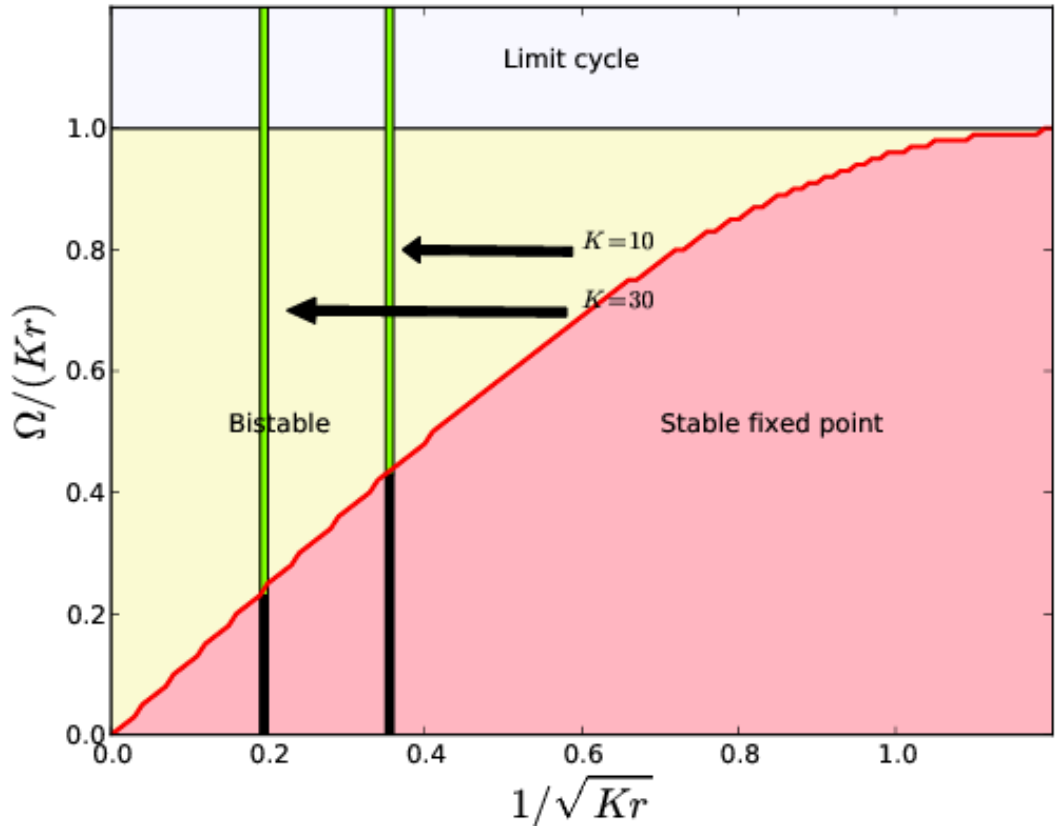}
\caption{(color online). The definitions of three shaded areas are the same as that in Fig.~\ref{Fig:natural_frequency_theta}. Two boundaries are compared between coupling strengths $10$ and $30$. If oscillators are in locked state with black color and with Chartreuse color otherwise. Increasing the coupling strength $K$ further, the vertical line moves to the left.}
\label{Fig:parameter_space}
\end{figure}

To investigate how the phase synchronization boundary changes with different coupling strengths, we project the Fig.~\ref{Fig:natural_frequency_theta} on the $I$-$\beta$ parameter space and color the oscillators according to their stationary states in the parameter space. A comparison between the dynamics with average degree $10$ and that with $30$ is shown in Fig.~\ref{Fig:parameter_space}. We can see that oscillators with the same coupling share the same $\beta$ axis and the diversity of $I$ is due to the distribution of the natural frequencies $\Omega$. All synchronized nodes are inside the synchronized area, which is at the right side of the line $I=4\beta/\pi$.   

Therefore, after substituting the boundaries of the synchronized natural frequencies $\left[\Omega_{\textrm{lower}},\Omega_{\textrm{upper}}\right]$ and the Cauchy-Lorentz distribution into the definition of the order parameter $r$,
\begin{equation}
r=\left.\int^{\Omega_{\textrm{upper}}}_{\Omega_{\textrm{lower}}} \cos{(\theta_s)} g(\Omega) d\Omega\right.,
\end{equation}
where $\theta_s$ denotes the synchronized oscillator $\sin{(\theta_s)}=I$
and performing some mathematical manipulations, we get
\begin{equation}
r=\frac{2}{\pi Kr}\left[\sqrt{1+(Kr)^{2}}\arctan\left(\frac{4\sqrt{Kr+(Kr)^{3}}}{\sqrt{(\pi Kr)^{2}-16Kr}}\right)-\arcsin\left(\frac{4}{\pi\sqrt{Kr}}\right)\right].
\label{r_self_consisent}
\end{equation}

\begin{figure}
\includegraphics[width=0.9\linewidth]{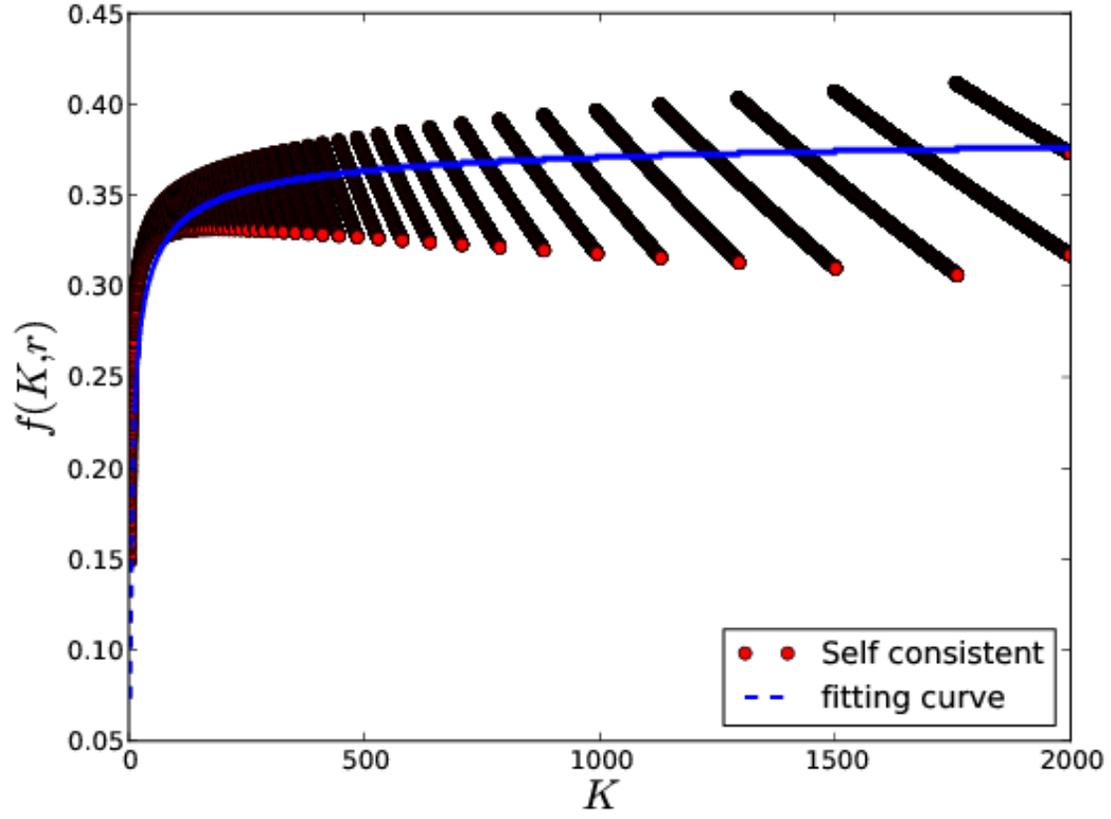}
\caption{(color online). $f(K,r)$ as a function of stationary solution of self-consistent equation colored in red and the fitting curve colored in blue. }
\label{Fig:high_order_term}
\end{figure}

Due to the difference of boundaries between the first-order Kuramoto model ( proportional to $K$) and the second-order Kuramoto model (proportional to $4\sqrt{K}/\pi$), we set $\Lambda(K)=4\sqrt{K}/\pi$.  When $\dot{r}=0$, 
\begin{equation}
f(K,r)=r-4\sqrt{K}r(1-r^2)/(2\pi),
\label{stationary_high_order}
\end{equation}
and this stationary solution should be met by the self-consistent Eq.~(\ref{r_self_consisent}). Here, we use numerical methods to calculate the values of $a$, $b$ and $c$. As shown in Fig.~\ref{Fig:high_order_term}, 
after substituting the stationary solutions $K$ and $r$ of Eq.~(\ref{r_self_consisent}) into Eq.~(\ref{stationary_high_order}), $f(K,r)$ is colored in red and we get the values $a=0.389$, $b=1/4$ and $c=3$. When $r$ is small, $f(K,r)$ is close to $0$ and cannot influence the time evolution of the order parameter $r(t)$, varying the stationary solution, otherwise. 

\begin{figure}
\includegraphics[width=0.9\linewidth]{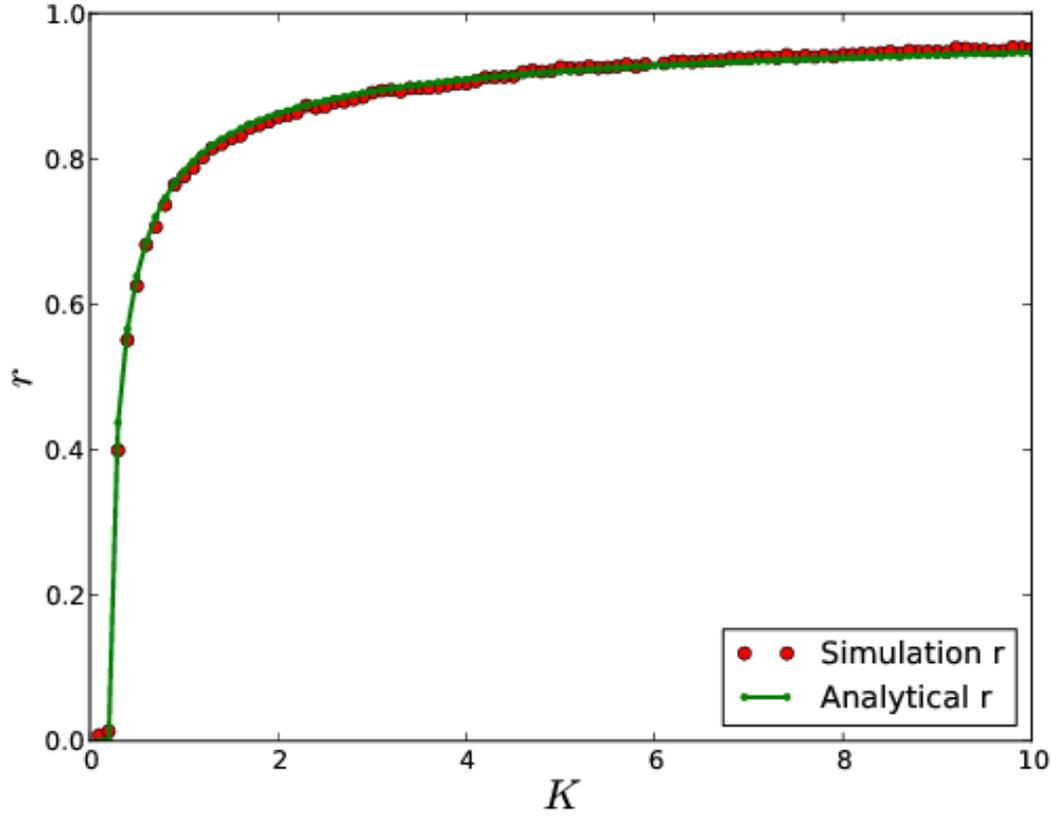}
%{data/r_stationary_K}
\caption{(color online). Order parameter $r$ vs coupling strengths $K$ in scale-free networks (see Methods for details). The red curves indicate the results from simulations on the same network as in fig.~\ref{Fig:figure_r_1st}. For each coupling, initial values of $\theta$ randomly select from $[-\pi,\pi]$ and we set $\dot{\theta}=0$. The green dots shows analytic prediction of the stationary $r(t)$ based on the self-consistent Eq.~(\ref{seek_nonlinear_time_evolution_second}).}
\label{Fig:r_stationary_K}
\end{figure}

Let us consider again the nonlinear evolution of the order parameter $r$ in complex networks. From the above analysis, we get that $\Lambda(Kk)=4\sqrt{Kk}/\pi$.
To check the validity of this assumption, we compare the stationary solution with simulation results in Fig.~\ref{Fig:r_stationary_K}. The theoretical predictions (green lines derived from Eq.~(~\ref{seek_nonlinear_time_evolution_second}) with effective coupling and $f(Kk,r)$) are in agreement with red lines of numerical simulations. 

\begin{figure}[!]
\includegraphics[width=0.9\linewidth]{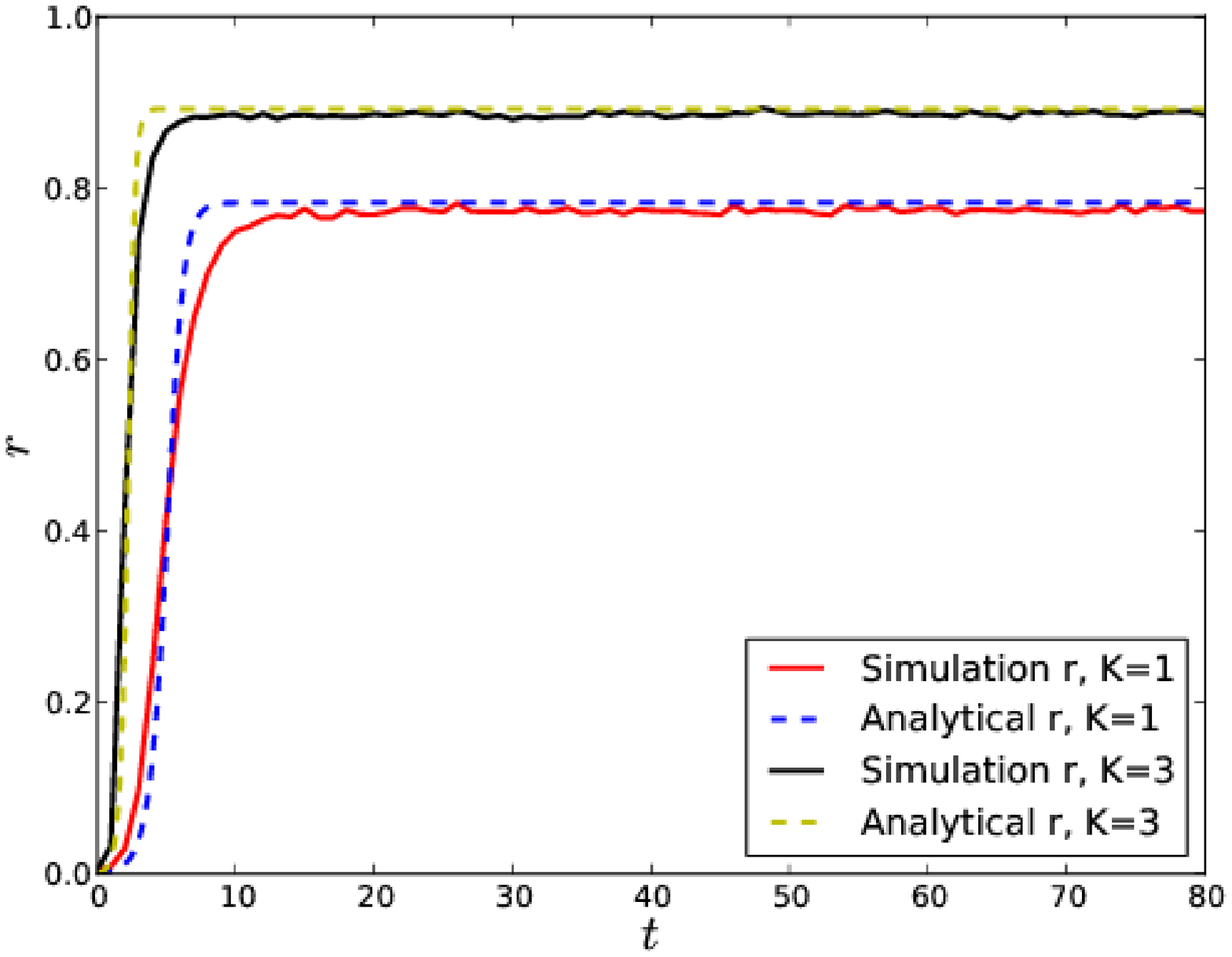}
%{data/r_ansatz_K_time}
\caption{(color online). Order parameter $r(t)$ vs time $t$ in scale-free networks (see Methods for details). The simulations are conducted on the same network and the coupling strength $K=1$ and $K=3$. Blue and yellow dots are analytic results got from Eq.~(\ref{seek_nonlinear_time_evolution_second}). In simulations, initial values of $\theta$ are randomly selected from $-\pi$ to $\pi$ and that of $\dot\theta$ close to $0$. } 
\label{Fig:r_t_ansatz}
\end{figure}

The nonlinear evolution of $r(t)$ is illustrated in Fig.~\ref{Fig:r_t_ansatz}, for a selection of coupling strengths $K$. Initial values of $\theta_i$ and $\dot{\theta}_i$ are the same as in Fig.~\ref{Fig:r_stationary_K}. For the order parameter formulation the initial value of $r$ is set to a small value $(r(0)\lll 1)$. The $r$ formulation of Eq.~(\ref{seek_nonlinear_time_evolution_second}) does not only reproduce the stationary states in Fig.~\ref{Fig:r_stationary_K}, but also matches the transition to synchrony. The analytic results and simulation results are in good agreement. 

\section*{Conclusions}

In conclusion, we proposed a generalization for the Ott-Antonsen ansatz to complex networks with a Cauchy-Lorentz distribution of the natural frequency for the Kuramoto model. Compared to the ensemble approach ~\cite{Barlev011}, the dimension of ordinary differential equations was reduced from $N$ to the number of possible degrees in the network.  We have investigated the collective dynamics of the Kuramoto model with inertia and found the synchronization boundary is $\left[-4\sqrt{Kr}/\pi,4\sqrt{Kr}/\pi\right]$ instead of $\left[-Kr,Kr\right]$ as in the Kurmoto model without inertia. Based on these results, we analytically derived self-consistent equations for the order parameter and nonlinear time-dependent order parameter. The agreement between the analytical and simulation results is excellent.

\section*{Methods}

\textbf{The networks}: 
The model has been implemented on undirected and unweighted scale-free networks with $N=10000$, $P(k)\propto k^{-3}$ and $k\geq5$.

\textbf{Numerical integration}:
Eqs.~(\ref{r_k_evolves}) and~(\ref{stationary_high_order}) are solved by a $4^{th}$ Rung-Kutta method with time step $h=0.01$ and with the Cauchy-Lorentz distribution $g(\Omega)=\frac{1}{\pi(1+\Omega^2)}$. 

\textbf{Order parameter}:
In complex networks, in order to understand the dynamics of the system, it is natural to use the definition of order parameter $r$~\cite{Ichinomiya04:PRE} as $r e^{i\psi}=\frac{\sum_i k_i e^{i\theta_i}}{\sum_i k_i}$ instead of the definition $r e^{i\psi}=\frac{\sum_i e^{i\theta_i}}{N}$, which accounts for the mean-field in the fully connected graph regime. 

The magnitude $r \in [0,1]$ quantifies the phase coherence, while $\psi$ denotes the average phase of the system. In particular, $r\simeq 0$, if the phases are randomly distributed over $[0,2\pi]$ and all nodes oscillate at its natural frequency. On the other hand, if all oscillators run as a giant component, $r \simeq 1$. The system is known to exhibit
a phase transition from the asynchronous state ($r \simeq 0$) to the synchronous one ($r\simeq 1$) at a certain critical
value $\lambda_c$ characterizing the onset of partial synchronization and, for unimodal and symmetric frequency distributions $g(\Omega)$, the transition is continuous. It turns out that for uncorrelated networks, $\lambda_c$ is given by $\lambda_c = \frac{2}{\pi g(\Omega) \lambda_{\max}}$~\cite{Restrepo05:PRE}, where $\lambda_{\max}$ is the maximal eigenvalue of the adjacency matrix. 

\section*{Acknowledgments}

P. Ji would like to acknowledge China Scholarship Council (CSC) scholarship. 
T. Peron would like to acknowledge FAPESP (No. 2012/22160-7) and within the scope of IRTG 1740.
F. A. Rodrigues acknowledge CNPq (grant 305940/2010-4), FAPESP (grant 2011/50761-2 and 2013/26416-9) and NAP eScience - PRP - USP for the financial support given to this research. 
J. Kurths would like to acknowledge IRTG 1740 (DFG and FAPESP) for the sponsorship provided.
P. Ji is very grateful to S. Petkoski, V. Kohar, Dr. Yanchuk and Dr. Stemler for many inspiring discussions. 

\section*{Author contributions}
P. Ji, T. Peron, F. A. Rodrigues. and J. Kurths designed and performed the research, analyzed the results and wrote the paper.

\section*{Additional information}
Competing financial interests: The authors declare no competing financial interests.
Correspondence and requests for materials should be addressed to  (e-mail: pengji@pik-potsdam.de, thomas.peron@usp.br, francisco@icmc.usp.br)

\newpage

% \begin{thebibliography}{10}
% 
% \end{thebibliography}

%\bibliographystyle{unsrt}
%\bibliography{paper}

\begin{thebibliography}{10}

\bibitem{Pikovsky03}
A.~Pikovsky, M.~Rosenblum, and J.~Kurths.
\newblock {\em Synchronization: A universal concept in nonlinear sciences},
  volume~12.
\newblock Cambridge University Press, 2003.

\bibitem{Arenas08:PR}
A.~Arenas, A.~D{\'\i}az-Guilera, J.~Kurths, Y.~Moreno, and C.~Zhou.
\newblock Synchronization in complex networks.
\newblock {\em Physics Reports}, 469(3):93--153, 2008.

\bibitem{buck1988synchronous}
John Buck.
\newblock Synchronous rhythmic flashing of fireflies. ii.
\newblock {\em Quarterly Review of Biology}, pages 265--289, 1988.

\bibitem{sherman1991model}
Arthur Sherman and John Rinzel.
\newblock Model for synchronization of pancreatic beta-cells by gap junction
  coupling.
\newblock {\em Biophysical journal}, 59(3):547--559, 1991.

\bibitem{schafer1998heartbeat}
Carsten Sch{\"a}fer, Michael~G Rosenblum, J{\"u}rgen Kurths, and Hans-Henning
  Abel.
\newblock Heartbeat synchronized with ventilation.
\newblock {\em Nature}, 392:239--240, 1998.

\bibitem{strogatz2005theoretical}
Steven~H Strogatz, Daniel~M Abrams, Allan McRobie, Bruno Eckhardt, and Edward
  Ott.
\newblock Theoretical mechanics: Crowd synchrony on the millennium bridge.
\newblock {\em Nature}, 438(7064):43--44, 2005.

\bibitem{Kuramoto_Yoshiki_lecture_notes_1975}
Yoshiki Kuramoto.
\newblock Self-entrainment of a population of coupled non-linear oscillators.
\newblock In Huzihiro Araki, editor, {\em International Symposium on
  Mathematical Problems in Theoretical Physics}, volume~39 of {\em Lecture
  Notes in Physics}, pages 420--422. Springer Berlin Heidelberg, 1975.

\bibitem{Ott_Antonsen_ansatz_chaos_2008}
Edward Ott and Thomas~M. Antonsen.
\newblock Low dimensional behavior of large systems of globally coupled
  oscillators.
\newblock {\em Chaos: An Interdisciplinary Journal of Nonlinear Science},
  18(3):--, 2008.

\bibitem{PhysRevE.86.046212}
Spase Petkoski and Aneta Stefanovska.
\newblock Kuramoto model with time-varying parameters.
\newblock {\em Phys. Rev. E}, 86:046212, Oct 2012.

\bibitem{PhysRevLett.103.044101}
Wai~Shing Lee, Edward Ott, and Thomas~M. Antonsen.
\newblock Large coupled oscillator systems with heterogeneous interaction
  delays.
\newblock {\em Phys. Rev. Lett.}, 103:044101, Jul 2009.

\bibitem{PhysRevLett.109.164101}
Oleh~E. Omel'chenko and Matthias Wolfrum.
\newblock Nonuniversal transitions to synchrony in the sakaguchi-kuramoto
  model.
\newblock {\em Phys. Rev. Lett.}, 109:164101, Oct 2012.

\bibitem{PhysRevLett.110.064101}
D.~Iatsenko, S.~Petkoski, P.~V.~E. McClintock, and A.~Stefanovska.
\newblock Stationary and traveling wave states of the kuramoto model with an
  arbitrary distribution of frequencies and coupling strengths.
\newblock {\em Phys. Rev. Lett.}, 110:064101, Feb 2013.

\bibitem{PhysRevE.88.012905}
Yi~Ming Lai and Mason~A. Porter.
\newblock Noise-induced synchronization, desynchronization, and clustering in
  globally coupled nonidentical oscillators.
\newblock {\em Phys. Rev. E}, 88:012905, Jul 2013.

\bibitem{Barlev011}
Gilad Barlev, Thomas~M. Antonsen, and Edward Ott.
\newblock The dynamics of network coupled phase oscillators: An ensemble
  approach.
\newblock {\em Chaos: An Interdisciplinary Journal of Nonlinear Science},
  21(2):--, 2011.

\bibitem{PhysRevLett.110.218701}
Peng Ji, Thomas K.~DM. Peron, Peter~J. Menck, Francisco~A. Rodrigues, and
  J\"urgen Kurths.
\newblock Cluster explosive synchronization in complex networks.
\newblock {\em Phys. Rev. Lett.}, 110:218701, May 2013.

\bibitem{strogatz1994nonlinear}
S.H. Strogatz.
\newblock {\em Nonlinear Dynamics and Chaos. With Applications to Physics,
  Chemistry and Engineering}.
\newblock Reading, PA: Addison-Wesley, 1994.

\bibitem{Acebron05:RMP}
J.~A. Acebr{\'o}n, L.~L. Bonilla, C.~J.~P. Vicente, F.~Ritort, and R.~Spigler.
\newblock The kuramoto model: A simple paradigm for synchronization phenomena.
\newblock {\em Rev. Mod. Phys.}, 77(1):137, 2005.

\bibitem{Dorfler14012013}
Florian D\"orfler, Michael Chertkov, and Francesco Bullo.
\newblock Synchronization in complex oscillator networks and smart grids.
\newblock {\em Proc. Natl. Acad. Sci. U.S.A}, 2013.

\bibitem{tanaka1997self}
H.A. Tanaka, A.J. Lichtenberg, and S.~Oishi.
\newblock Self-synchronization of coupled oscillators with hysteretic
  responses.
\newblock {\em Physica D: Nonlinear Phenomena}, 100(3):279--300, 1997.

\bibitem{sonnenschein2013approximate}
Bernard Sonnenschein and Lutz Schimansky-Geier.
\newblock Approximate solution to the stochastic kuramoto model.
\newblock {\em Physical Review E}, 88(5):052111, 2013.

\bibitem{Ichinomiya04:PRE}
Takashi Ichinomiya.
\newblock Frequency synchronization in a random oscillator network.
\newblock {\em Phys. Rev. E}, 70:026116, 2004.

\bibitem{PhysRevE.86.056108}
Thomas Kau\^e~Dal'Maso Peron and Francisco~A. Rodrigues.
\newblock Determination of the critical coupling of explosive synchronization
  transitions in scale-free networks by mean-field approximations.
\newblock {\em Phys. Rev. E}, 86:056108, Nov 2012.

\bibitem{strogatz2000kuramoto}
S.H. Strogatz.
\newblock From kuramoto to crawford: exploring the onset of synchronization in
  populations of coupled oscillators.
\newblock {\em Physica D: Nonlinear Phenomena}, 143(1):1--20, 2000.

\bibitem{Barlev_2011_chaos_an_ensemble_approach}
Gilad Barlev, Thomas~M. Antonsen, and Edward Ott.
\newblock The dynamics of network coupled phase oscillators: An ensemble
  approach.
\newblock {\em Chaos: An Interdisciplinary Journal of Nonlinear Science},
  21(2):--, 2011.

\bibitem{PhysRevLett.107.178701}
Charo~I. Del~Genio, Thilo Gross, and Kevin~E. Bassler.
\newblock All scale-free networks are sparse.
\newblock {\em Phys. Rev. Lett.}, 107:178701, Oct 2011.

\bibitem{guckenheimer1983nonlinear}
J.~Guckenheimer and P.~Holmes.
\newblock {\em Nonlinear oscillations, dynamical systems, and bifurcations of
  vector fields}, volume~42.
\newblock Springer-Verlag New York, 1983.

\bibitem{Restrepo05:PRE}
Juan~G. Restrepo, Edward Ott, and Brian~R. Hunt.
\newblock Onset of synchronization in large networks of coupled oscillators.
\newblock {\em Phys. Rev. E}, 71:036151, 2005.

\end{thebibliography}

\end{document}